\documentstyle[epsfig]{aipproc}

\begin{document}
\title{Rare Kaon Decays}

\author{Toshio Numao}
\address{TRIUMF, 4004 Wesbrook Mall, Vancouver, B.C.,
Canada V6T~2A3\\
email: toshio@triumf.ca}

\maketitle

\begin{abstract}
Rare kaon decays via Flavor Changing Neutral Currents
are discussed in the context of the CKM unitarity triangle
with a particular interest in the
rare kaon decays $K^+ \rightarrow \pi^+ \nu \bar{\nu}$
and $K^0_L \rightarrow \pi^0 \nu \bar{\nu}$.
New results and the status of these experiments are reported.
\end{abstract}

\section*{Introduction}

The study of
rare kaon decays has a glorious history.
At the level of $10^{-1}$ in branching ratio,
parity violation was first ``discovered''
as the $\theta$-$\tau$ puzzle\cite{leeyang},
CP violation was discovered at
the level of $10^{-3}$\cite{firstcp}, and the
GIM mechanism\cite{gim} was suggested by the strong
suppression of the decay
$K^0_L \rightarrow \mu^+ \mu^-$, or the absence of
the Flavor Changing Neutral Currents (FCNC)---
the decay mode was subsequently found
at the level of $10^{-8}$. Further
searches for rare kaon decays at lower branching ratio
undoubtably involve rich physics and may yet offer another surprise---
e.g. they are likely to
elucidate the origin of CP violation, and provide
some clues for physics beyond the Standard Model (SM).
In this talk,
the decay modes via FCNC with
the expected branching ratios of less than
$10^{-9}$ are discussed.
Since exotic decays violating the lepton flavor
conservation law
have been discussed by the previous speaker\cite{klaus},
the focus of
this talk is on the rare kaon decays which are allowed in the SM.

\begin{figure}[b!] 
\centerline{\epsfig{file=fig1.epsi,height=1.5in,width=5.0in}}
\vspace{10pt}
\caption{Typical Feynman diagrams.}
\label{fig1}
\end{figure}

Typical leading Feynman diagrams of these decays are shown in Fig.~1,
where $d$ and $u$ at the top lines are for $K^0$ and $K^+$
decays, respectively, and
$\ell$ indicates the electron or the muon.
Because of the GIM mechanism, the lowest order diagrams come from
second order weak interactions with a $u$-type quark
in the loop diagrams,
in which the top-quark contributions dominate
because of the mass.
This makes
these decay modes very sensitive to $V_{td}$,
the least constrained coupling-constant of the top and down quarks in
the Cabibbo-Kobayashi-Maskawa (CKM) matrix.
Due to one of the unitarity conditions
of the CKM matrix
$V_{ud}V^*_{ub}+V_{cd}V^*_{cb}+V_{td}V^*_{tb}=0$
and with the approximation
$V_{ud} \sim V_{tb} \sim 1$,
$V_{td}$ and $V^*_{ub}$ form
two sides of a unitarity triangle  as shown in Fig.~2.
The height of the triangle
$Im(V_{td})$ is an indication of ``direct'' CP violation,
or CP violation through the decay amplitude.
For the decay $K^+ \rightarrow \pi^+ \nu \bar{\nu}$,
the branching ratio is roughly proportional to $|V_{td}|^2$
and for $K^0_L \rightarrow \pi^0 \nu \bar{\nu}$ to $Im^2 (V_{td})$.
These measurements alone can determine the unitarity triangle,
and when they are combined with measurements from $B$-decays
it will provide over-constrained information that is
sensitive to a presence of supersymmetry and other physics
beyond the SM\cite{bb,nir}.

\begin{figure}[b!] 
\centerline{\epsfig{file=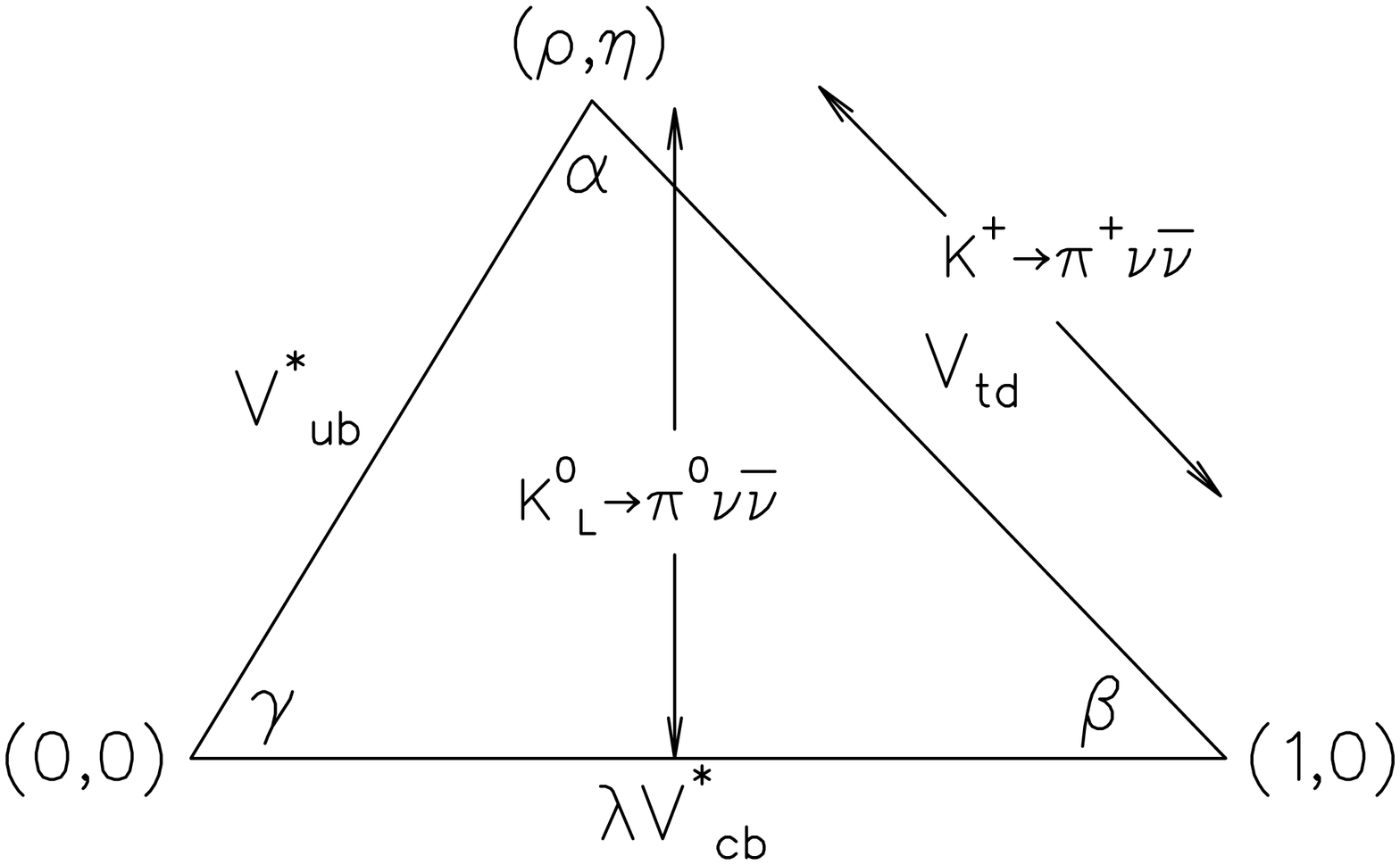,height=4.0in,width=2.5in,
angle=90}}
\vspace{10pt}
\caption{Unitarity triangle.
The coordinates of the
vertices are for the rescaled triangle[8].}
\label{fig2}
\end{figure}

\section*{Decays
$K^0_L \rightarrow \pi^0 \ell^+ \ell^-$ and $K^0_L \rightarrow
\ell^+ \ell^-$}

If the final states include a charged lepton pair, there are
additional contributions from diagrams with virtual photons
that usually prevent clear interpretations of measurements\cite{valencia}.

The decays $K^0_L \rightarrow \pi^0 \ell^+ \ell^-$ have a
``direct'' CP violating component, which is expected to
occur, if it were the only component, at a branching ratio of
$\sim 5 \times 10^{-12}$. There are two other contributions,
however, at the same level to this
process; one arises from the mixing of the CP even state in
$K^0_L$, and the other from two-virtual-photon intermediate states
that conserve CP. At present,
there are theoretical ambiguities in the estimations
of these contributions, which
may eventually be sorted out
by measurements of $K^0_S \rightarrow \pi^0 \ell^+ \ell^-$ and
other radiative decays.
To make the matter a bit more complicated,
there is a physical background coming from the radiative decay
$K^0_L \rightarrow \ell^+ \ell^- \gamma \gamma$,
which has the identical event topology.
With the tightest
cuts, the background level is estimated to be still at a $10^{-11}$ level
for $\ell=e$\cite{greenlee}.
The situation is similar in the case of $\ell=\mu$.
The present upper limits (90 \% c.l.) are
$B(K^0_L \rightarrow \pi^0 e^+ e^-) \leq 5.6 \times
10^{-10}$\cite{pee} and
$B(K^0_L \rightarrow \pi^0 \mu^+ \mu^-)
\leq 3.8 \times 10^{-10}$\cite{pmm}.

The decay $K^0_L \rightarrow e^+ e^-$ is very similar to
$K^0_L \rightarrow \mu^+ \mu^-$, which is sensitive to the $\rho$
parameter\cite{lw}, but it is further suppressed by
the helicity mechanism. This decay is sensitive to pseudo-scalar
interactions coming from physics beyond the SM.
Four events from the decay $K^0_L \rightarrow e^+ e^-$ have been
reported recently which correspond to a branching ratio,
$B(K^0_L \rightarrow e^- e^+) =
8.7^{+5.7}_{-4.1} \times 10^{-12}$\cite{ee}, being consistent
with the unitarity bound expected from the long-distance
contributions.\\

\section*{Decay $K^+ \rightarrow \pi^+ \nu \bar{\nu}$}

In the SM calculation of the decay
$K^+ \rightarrow \pi^+ \nu \bar{\nu}$, the dominant contribution
comes from second order loop diagrams with a virtual top quark
(Fig.~1).
The hadronic matrix element can be extracted from the decay
$K \rightarrow \pi^0 e^+ \nu$ and
theoretical uncertainties in the calculation due to long distance
contributions and other effects are small\cite{bb}.
The signature of the decay $K^+ \rightarrow \pi^+ \nu \bar{\nu}$
is a single pion with no other observable decay-products.
Definitive observation of this signal requires that all possible
backgrounds are suppressed well below
the signal level.
Major background sources are:
a muon from the copious decay $K^+ \rightarrow \mu^+ \nu$ ($K_{\mu 2}$)
which is misidentified as a pion,
a pion from the decay $K^+ \rightarrow \pi^+ \pi^0$ ($K_{\pi 2}$)
when two photons from the $\pi^0$ decay are unobserved,
a beam pion scattered by the target into the detector,
and charge exchange reactions of $K^+$'s which result in
decays $K_L^0 \rightarrow \pi^+ \ell^- \bar{\nu}$,
where $\ell$  ($e$ or $\mu$) is undetected.
In order to suppress the backgrounds,
good particle identification, efficient photon veto,
and detection of incoming particles are essential.

\begin{figure}[b!] 
\centerline{\epsfig{file=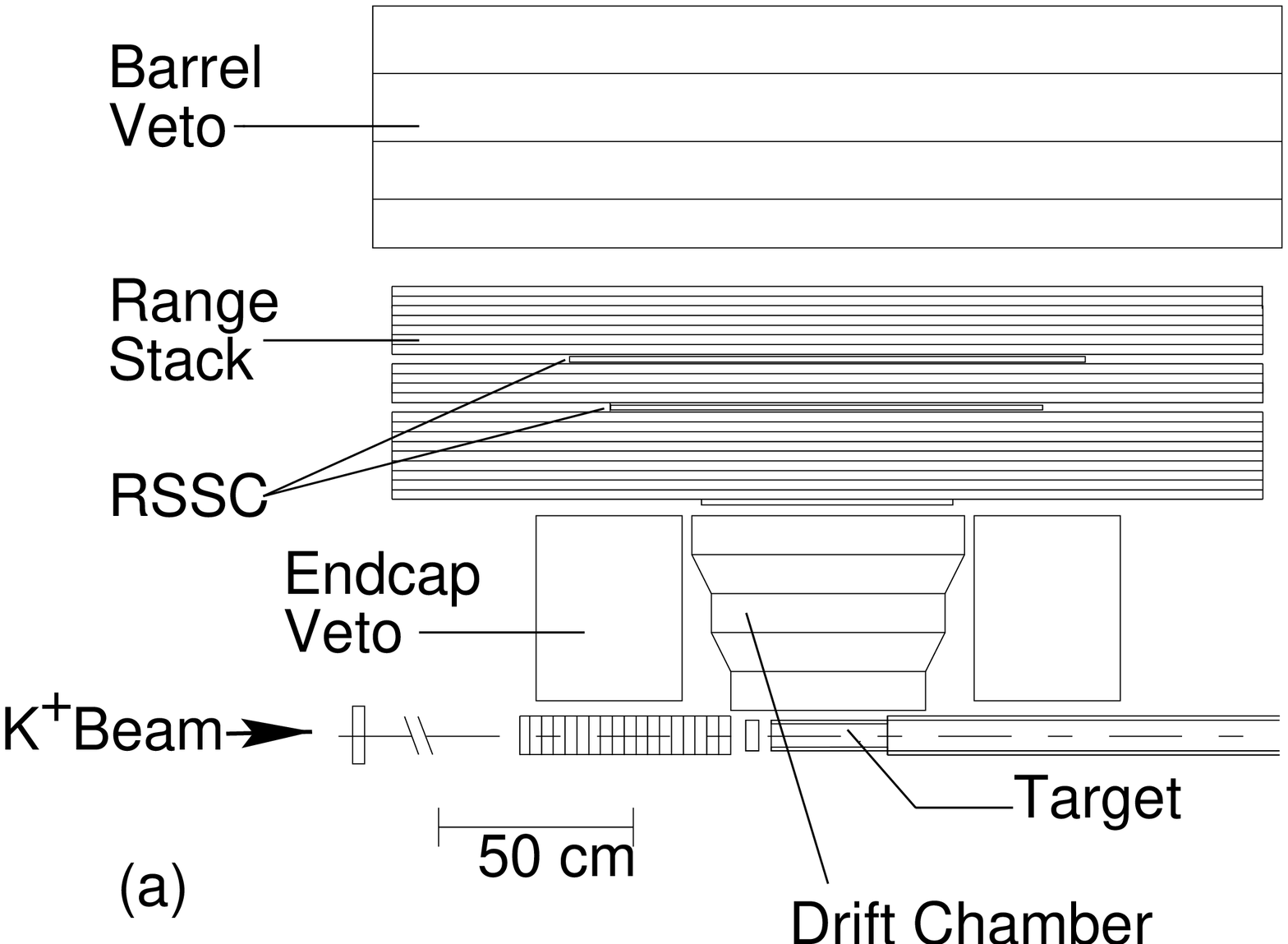,height=2.2in,width=2.7in}
\epsfig{file=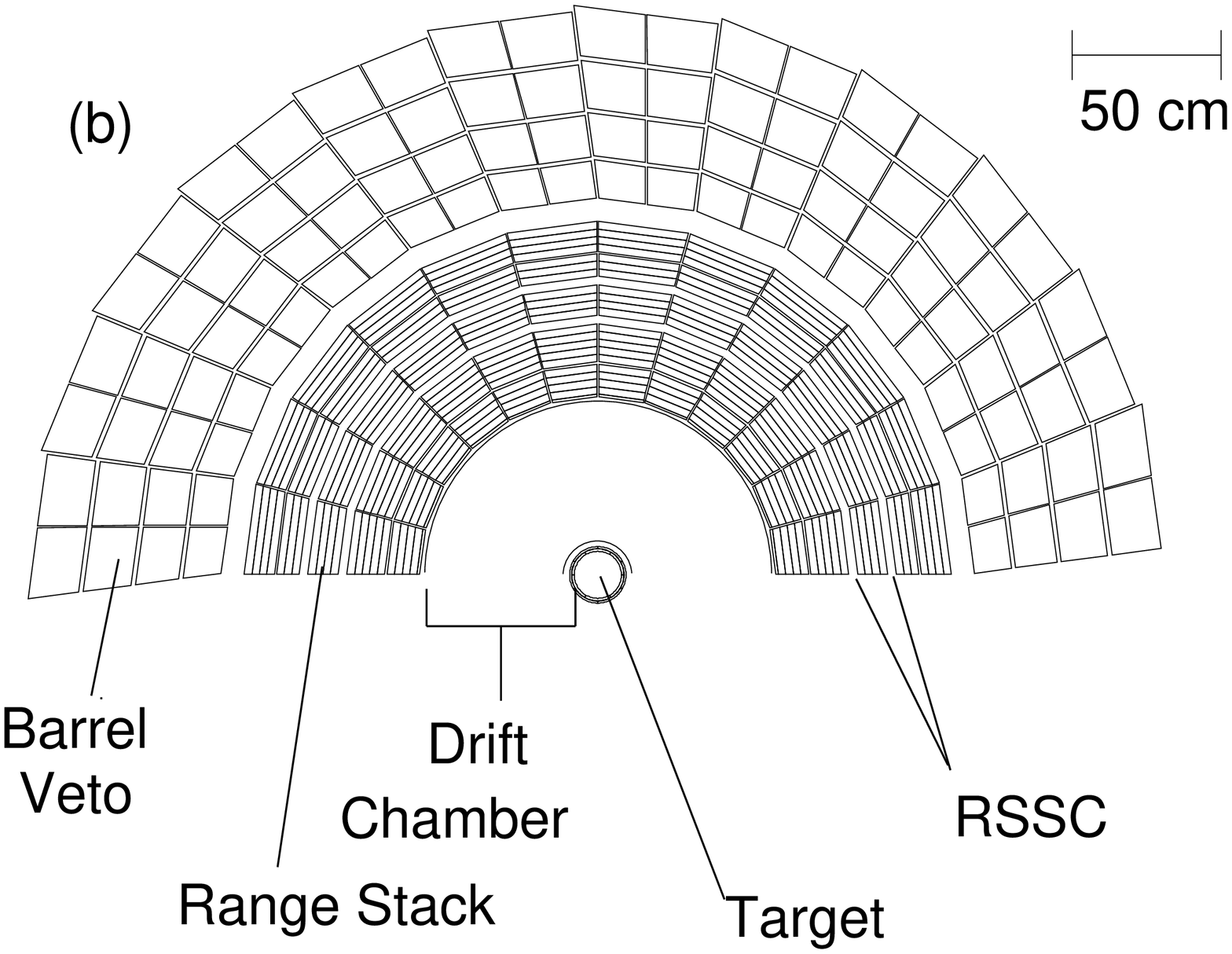,height=2.5in,width=2.7in}}
\vspace{10pt}
\caption{Upper half of the BNL E787/949 detector: (a) side view
and (b) end view.}
\label{fig3}
\end{figure}

The E787 experiment at Brookhaven National Laboratory (BNL)
as shown in Fig.~3
is designed to effectively distinguish these backgrounds
from the signal.
Kaons of about 700~MeV/c at a rate of
$(4-7) \times 10^6$ per 1.6-s spill are detected and identified
by a \v{C}erenkov counter and hodoscopes, degraded by BeO and
stopped in an active target, primarily consisting of
413 5-mm square scintillating fibers.
The momentum ($P$), kinetic energy ($E$) and range ($R$)
of decay products are measured using the target,
a central drift chamber, 21 layers of 1.9-cm thick plastic
scintillator (Range Stack) and two layers of straw chambers, all
contained in a 1-T magnetic field.
The $\pi^+ \rightarrow \mu^+ \rightarrow e^+$ decay sequence
of the decay products
in the Range Stack scintillator is observed
by 500-MHz transient digitizers for particle identification.
Photons are detected by a $4\pi$-sr calorimeter consisting
of a 14-radiation-length-thick lead/scintillator barrel detector,
13.4-radiation-length-thick end caps of CsI crystals,
and a 3.5-radiation-length-thick lead-glass
\v{C}erenkov counter which also works as an
active beam degrader.

The E787 experiment
reported an observation of one clean event at a branching ratio
$B(K^+ \rightarrow \pi^+ \nu \bar{\nu})
= 4.2^{+9.7}_{-3.5} \times 10^{-10}$ 
using the data sample taken in 1995\cite{e787a}.
Since then, additional data samples taken in 1996 and 1997,
together with those taken in 1995,
have been analyzed with improved algorithms, which have
resulted in
less non-Gaussian tails in $P$, $R$ and $E$ measurements, and
a $\sim$30 \% higher acceptance than that in Ref.\cite{e787a}.
Also, in the 1996--7 runs, lowering the incident $K^+$ beam momentum
resulted
in a larger fraction of kaons stopping in the target, which
reduced accidental hits originated from nuclear reactions in the
beam degrader. The higher proton intensity at
the production target compensated the kaon yield loss
at lower momentum.

In order to avoid a possible bias in the analysis,
the signal region is kept untouched until the background estimations
as well as optimization of acceptance have been done.
In the background study,
the data sample
after applying all the final cuts except two orthogonal (uncorrelated)
cut groups to be studied, e.g. the kinematical cuts and those related
to the observation of the decay sequence
$\pi^+ \rightarrow \mu^+ \rightarrow e^+$ in the stopping counter
for the estimation of the
$K^+ \rightarrow \mu^+ \nu$ background, is used to obtain
the suppression factor for each cut group.
The correlation between the two cut groups,
which may invalidate the method, is studied by varying
the cuts being studied or by enhancing certain types of background
events.

\begin{figure}[b!] 
\centerline{\epsfig{file=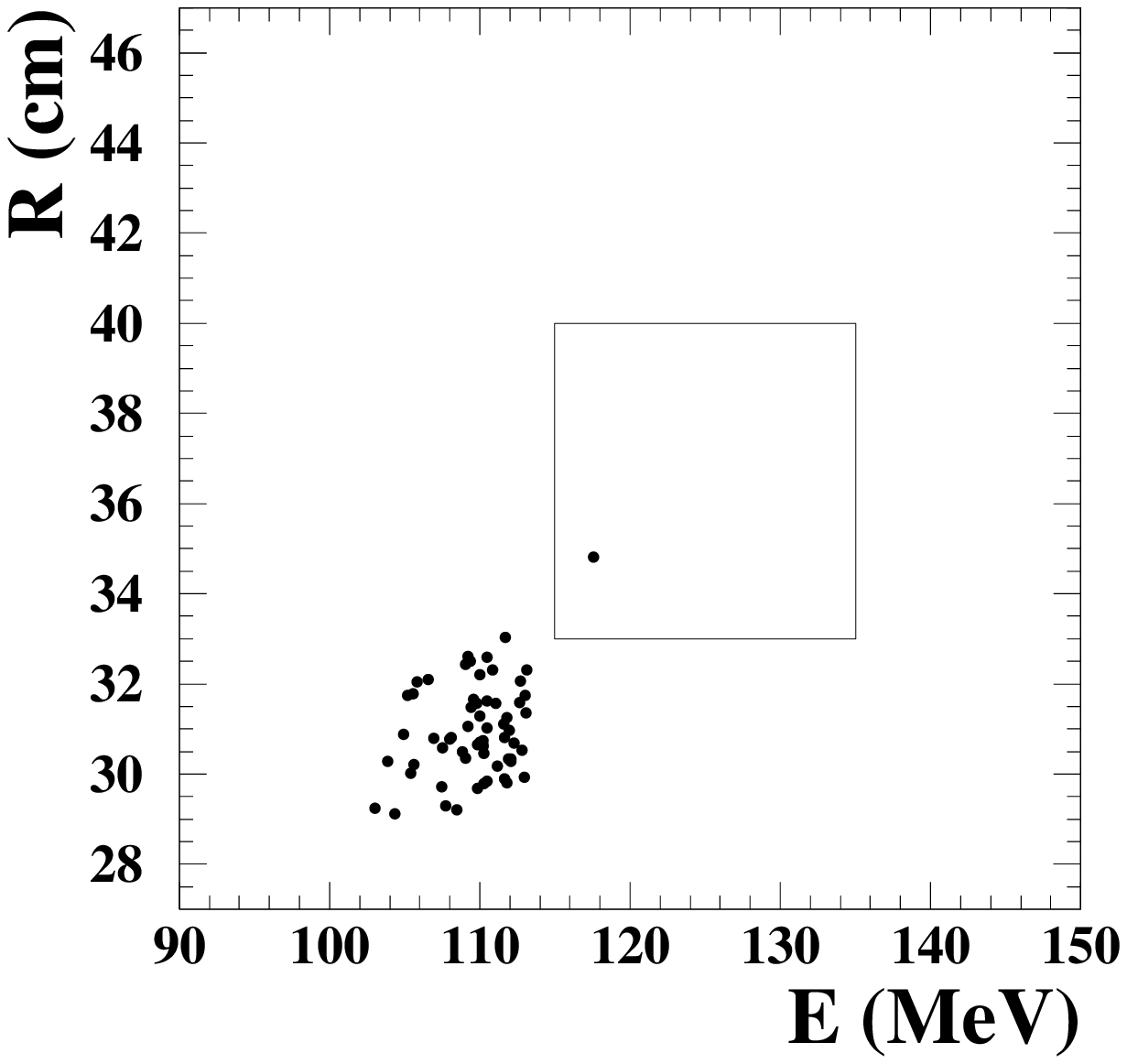,height=3.in,width=3.in}}
\vspace{10pt}
\caption{Range (cm in plastic scintillator) vs Kinetic energy
plot of the 95--97 data. The concentration at E=107 MeV is due to
$K_{\pi 2}$ events.}
\label{fig4}
\end{figure}

The total background for the entire 1995--1997 exposure
with the final analysis cuts is estimated to be $0.08 \pm 0.02$
events.
The acceptance for $K^+ \rightarrow \pi^+ \nu \bar{\nu}$,
$A = 0.0021 \pm 0.0001(stat) \pm 0.0002(syst)$ is calculated
based on data and Monte Carlo calculations. The largest uncertainty
comes from the uncertainty in pion-nucleus interaction.
The measurement of
the branching ratio for $K^+ \rightarrow \pi^+ \pi^0$ within  a few \%
of Ref.\cite{pdg} confirms the acceptance calculation.
Analysis of the full data sample  as shown in Fig.~4
has yielded only the single event
previously reported.
Based on the acceptance $A$ and the total exposure of
$N_{K^+} = 3.2 \times 10^{12}$ kaons, the new branching ratio
is $B(K^+ \rightarrow \pi^+ \nu \bar{\nu}) =
1.5^{+3.4}_{-1.2} \times 10^{-10}$\cite{new}.
This provides a constraint, $0.002 < |V_{td}| < 0.04$.

The goal of a new experiment E949 at BNL is to improve
the sensitivity by an order of magnitude.
Since the AGS in the RHIC era will be used for two hours a day
to feed heavy ions into the RHIC ring,
the rest of 22 hours can be used for the high energy program.
The operation is expected to provide a more stable and longer
running period.
Exploiting a higher beam intensity,
the incident $K^+$ momentum can be further lowered to reduce
accidental coincidence.
The photon veto capability will be improved by additional
lead/scintillator layers in the barrel region and by
additional active degrader in the beam region.
In the region below the $K^+ \rightarrow \pi^+ \pi^0$
peak at 205~MeV/c, where nuclear interactions result in a
large momentum-tail,
the additional photon veto capability may suppress
the background by more than an order of magnitude
as low as to the signal level, doubling the phase space in the
search.

\subsection*{Decay $K^0_L \rightarrow \pi^0 \nu \bar{\nu}$}

The decay $K^0_L \rightarrow \pi^0 \nu \bar{\nu}$
violates  the CP conservation law through decay amplitude.
This process is expected to occur at
$B(K^0_L \rightarrow \pi^0 \nu \bar{\nu}) \sim 3 \times 10^{-11}$
\cite{bb}.
The contribution from CP mixing in $K_L^0$ is expected to be
around $10^{-15}$\cite{ll}. Since no charged leptons are involved
in this decay, the process is free from virtual photon contributions
and clean for the study of the origin of CP violation.
The theoretical ambiguity is only $\sim 1~\%$ except those in the
CKM matrix elements.
Conversely, the observation of this decay mode
uniquely determines $Im(V_{td})$ or $\eta$ in the Wolfenstein
parametrization.
The goal of the experiment is to determine $Im(V_{td})$ with a
10--15 \% accuracy, which corresponds to a single event sensitivity
of $10^{-12}$.
The present upper limit of this decay is $5.7 \times 10^{-7}$\cite{ktev}.

The decay $K^0_L \rightarrow \pi^0 \nu \bar{\nu}$
is a three-body decay that involves only neutral particles.
The signature of this decay is two photons from the $\pi^0$ and no
other activity in the detector. In this decay mode,
available kinematical parameters
are very limited; relatively
easy ones to measure are the positions and energies
of $\gamma$-rays.
Also, the direction and position of the $K_L^0$  beam can be
limited by tightly collimating the beam at a cost of beam
intensity.
The decay vertex can be reconstructed from
these two constraints with the assumption of the pion mass.
The KEK experiment\cite{inagaki} and the FNAL approach\cite{kami}
are classified in this category.

The BNL experiment E926\cite{e926}
attempts to measure more kinematic values;
the directions
of $\gamma$-rays and the momentum of the $K_L^0$.
Measurements of $\gamma$-ray directions
allow full reconstruction of the $\pi^0$ kinematics
without the assumption of the $\pi^0$ mass,
providing more constraints
with redundancy,
which is necessary to suppress the background to the level
well below the signal.
The time-of-flight (TOF) of a $K_L^0$
between the production target and the decay vertex
can be measured for low momentum
$K_L^0$'s if the incident proton beam is bunched. This measurement
allows calculation of the missing mass and
kinematical reconstruction
in the center of mass system, which is effective to eliminate
backgrounds from two-body decays.
The major decay modes of $K_L^0$ are
$K_L^0 \rightarrow \pi^0 \pi \pi$ and
$K_L^0 \rightarrow \pi^{\pm} \ell^{\mp} \nu$.
Suppression of most backgrounds is achieved by high-efficiency
hermetic photon and charged-particle detector system
surrounding the decay volume, and kinematical constraints.

\begin{figure}[b!] 
\centerline{\epsfig{file=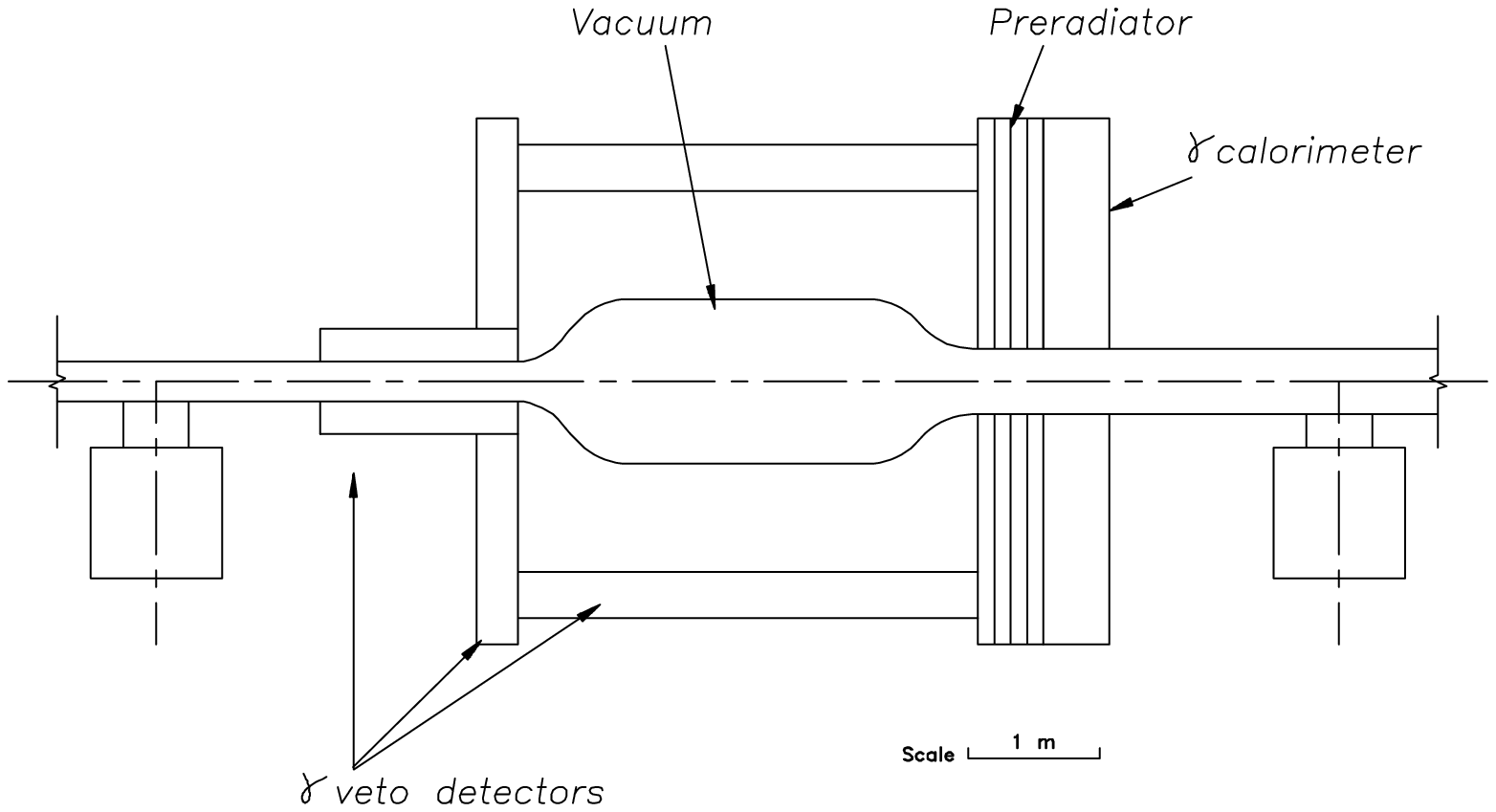,height=2.7in,width=4.3in}}
\vspace{10pt}
\caption{A typical detector system for
$K_L^0 \rightarrow \pi^0 \nu \bar{\nu}$ experiments.}
\label{fig5}
\end{figure}

In the BNL E926 experiment,
a low energy $K_L^0$ beam with an average momentum around 650~MeV/c
is produced by irradiating a target
with a 24-GeV proton beam from the AGS
and extracted  at 40$^o$ with respect to the incident beam.
The proton beam is bunched to form
a $\leq$200~ps wide bucket at a rate of 25~MHz.
About 16 \% of $K_L^0$'s decay in the 4-m long decay volume,
which is evacuated to a level of $10^{-7}$~Torr to suppress
the background from neutron-induced $\pi^0$ production.
The decay region is surrounded by a charged particle veto system
and a photon veto system of 18-radiation-length-thick
lead/scintillator sandwiches in the barrel region.
The two photons from the $\pi^0$ decay are converted into  pairs of
a positron and an electron in a 2-radiation-length
preradiator next to the vacuum
region for the measurement
of the directions of the $\gamma$-rays.
The preradiator
consists of sandwiches of
2-mm thick scintillator, a copper plate as
a mechanical support and radiator, and a tracking chamber.
This is followed by an 18-radiation-length calorimeter for
the measurement of $\gamma$-ray energies and for vetoing
additional photons.

The estimates of sensitivity for $K_L^0 \rightarrow \pi^0 \nu
\bar{\nu}$
are tightly coupled to the cuts required for background suppression,
particularly for the
$K_L^0 \rightarrow \pi^0 \pi^0$ and
$K_L^0 \rightarrow \pi^0 \pi^+ \pi^-$ backgrounds.
An acceptance of $\sim 0.015$~\% for the case S/N=0.5
comes from the combination of factors; 0.58 for the fiducial region
and usable kaon momentum region, 0.33 for the solid angle, 0.5 for
the efficiency of the preradiator, and the remaining factor
for the $\pi^0$ mass cut and other cuts.
Assuming three years of running with $6 \times 10^6$ kaons per
2-s beam spill (50 \% duty factor), the expected number of
events is 65.

The toughest background is the CP violating
$K_L^0 \rightarrow \pi^0 \pi^0$ decay when two photons are undetected.
Backgrounds from $K_L^0 \rightarrow \pi^0 \pi^0$ arise when
two photons are detected in the forward detector and the other
two are undetected anywhere.
These backgrounds
can be classified into two categories; even pairing when two photons
come from the same $\pi^0$,
and odd pairing
when two photons
come from different $\pi^0$'s.
Events in the even pairing category form
a two-body-decay peak in the momentum spectrum of $\pi^0$
in the CM system, while
the two $\gamma$-rays in the odd pairing category do
not reproduce
the $\pi^0$ mass.
The above requirements and photon veto essentially suppress
$K_L^0 \rightarrow \pi^0 \pi^0$ backgrounds to 0.2 of
the signal level.
Similarly, $K_L^0 \rightarrow \pi^0 \pi^+ \pi^-$ backgrounds can
be suppressed to 0.1 of the expected signal.
Backgrounds from other $K_L^0$ decay modes are estimated to be
less than 0.1 of the signal.
Because of the large angle extraction, the cross sections for
producing $\Lambda$'s are small and they completely decay 
before reaching the decay volume. Backgrounds could arise from
$\Lambda$'s produced by halo neutrons and $K_L^0$'s, but they
are estimated to be negligible.
Neutrons with $P_n \geq 800$~MeV/c
can react on the remaining atoms in the vacuum region
to produce $\pi^0$'s. This background level is again estimated
to be 0.01.
Accidental backgrounds are caused by beam halo neutrons and
$\gamma$-rays which create a $\pi^0$ signal in the preradiator.
The background level is estimated to be 0.02 of the expected
signal.

The BNL E926 was proposed in 1997 but the group is still waiting for
full funding. In the present scenario,
the detector construction is expected to start in 2002.
The KEK experiment has partially been approved, but the
sensitivity goal is only around the SM level---the new
Japan Hadron Facility is expected to improve the sensitivity.

\section*{Conclusion}

Studies of rare kaon decays have contributed to the discoveries of
several symmetry violations.
The decays 
$K^+ \rightarrow \pi^+ \nu \bar{\nu}$ and
$K^0_L \rightarrow \pi^0 \nu \bar{\nu}$
are expected to ``complete'' the measurement of the CKM matrix
in the next decade independently from the $B$-decay system,
and to elucidate the origin of CP violation.

\end{document}